\documentclass{jaa}
\usepackage{graphicx}
\usepackage{caption}
\usepackage{subcaption}
\usepackage{hyperref}
\usepackage{array}

\begin{document}\sloppy

\title{Pulsars in AstroSat-CZTI: Detection in sub-MeV bands \and Estimation of Spectral Index from Hardness Ratios} 

\author{Anusree K.G.\textsuperscript{1,*}, Dipankar Bhattacharya\textsuperscript{2,5},  Varun Bhalerao\textsuperscript{3} and Akash Anumarlapudi\textsuperscript{4}}
\affilOne{\textsuperscript{1}School of Pure and Applied Physics, Mahatma Gandhi University, Kottayam 686560, India.\\}
\affilTwo{\textsuperscript{2}Inter-University Centre for Astronomy and Astrophysics, Pune 411007, India.\\}
\affilThree{\textsuperscript{3}Indian Institute of Technology, Bombay 400076, India.\\}
\affilFour{\textsuperscript{4}Department of Physics, University of Wisconsin-Milwaukee, Milwaukee, WI, USA \\}
\affilFive{\textsuperscript{5}Ashoka University, Sonipat 131029, India}
\twocolumn[{
\maketitle
\corres{anusree.kgk@gmail.com}

\begin{abstract}
    The Cadmium Zinc Telluride Imager (CZTI) onboard AstroSat,  an open detector above $\sim$100~keV, is a promising tool for the investigation of hard X-ray characteristics of $\gamma$-ray pulsars. A custom algorithm has been developed to detect pulsars from long integration ($\sim$years) of archival data, as reported by us earlier. Here we extend this method to include in the analysis additional $\sim$20\% of the CZTI pixels that were earlier ignored due to their lower gain values. Recent efforts have provided better and more secure calibration of these pixels, demonstrating their higher thresholds and extended energy range up to $\sim$1~MeV. Here we use the additional information provided by these pixels, enabling the construction of pulse profiles over a larger energy range. We compare the profiles of the Crab pulsar at different sub-bands and show that the behaviour is consistent with the extended energy coverage. As detailed spectroscopy over this full band remains difficult due to the limited count rate, we construct hardness ratios which, together with AstroSat Mass Model simulations, are able to constrain the power-law index of the radiation spectrum.  We present our results for the phase-resolved spectrum of PSR J0534+2200 and for the total pulsed emission of PSR J1513-5908. The recovered photon indices are found to be accurate to within $\sim 20$\%.
\end{abstract}
}]

\doinum{12.3456/s78910-011-012-3}
\artcitid{\#\#\#\#}
\volnum{000}
\year{0000}
\pgrange{1--}
\setcounter{page}{1}
\lp{1}


\section{Introduction}
    
    Rotation powered pulsars usually emit broadband covering radio through $\gamma$-rays. The geometry of the emission region and the radiation mechanism in the pulsar magnetosphere shape the pulse profile. However, the nature and origin of the non-thermal high energy emission still remains unresolved despite a large variety of theoretical models proposed over the years (Arons and Scharlemann 1979; Cheng, Ho and Ruderman 1986a,b; Romani and Yadigaroglu 1995; Kirk, Skjaeraasen and Gallant 2002; Dyks and Rudak 2003; Bai and Spitkovsky 2010; P\'{e}tri 2012; Philippov and Spitkovsky 2018 among others). One reason for the slow progress in this field has been the lack of detailed broadband observations of a significant population of pulsars at X-ray and $\gamma$-ray energies. The Fermi mission has been a game changer in the $\gamma$-ray band, increasing the number of known pulsars from seven to over 270, and this continues to grow\footnote{https://confluence.slac.stanford.edu/display/GLAMCOG/Public+List+of+LAT-Detected+Gamma-Ray+Pulsars}.  
    
   By modelling radio, X-ray, and $\gamma$-ray pulse profiles, important constraints on the magnetospheric properties of pulsars can be obtained. (e.g.\ Seyffert et al. 2011; Pierbattista et al. 2015; Lockhart et al. 2019; P\'{e}tri \& Mitra 2020). However, many of the Fermi-LAT $\gamma$-ray pulsars have no radio counterpart. PSR J0633+1746 (Abdo et al. 2010), PSR J2021+4026 (Lin et al. 2013) and PSR J1836+5925 (Abdo et al. 2013) are some examples. For these radio-quiet pulsars, the magnetospheric X-ray emission becomes the only possible source of additional information. Unfortunately, instruments capable of studying these pulsars in hard X-ray/soft $\gamma$-ray band have been rare. Most instruments with high timing and spectral resolution, that can conduct quality observations of known pulsars in this band, would still have to contend with the low flux levels of the majority of these pulsars, requiring prohibitively long pointed observations.  Such long integrations, however, become more feasible with open all-sky detectors where the target objects are always present in the field of view.
    
    The Cadmium Zinc Telluride Imager (hereafter CZTI, Bhalerao et al. 2017) onboard India's first Astronomical Satelite AstroSat (Singh et al. 2014; Paul 2013) becomes an open detector above ~100 KeV.  We have developed a
  custom algorithm that utilizes this property to detect pulsars, thus turning CZTI into a promising hard X-ray pulsar monitor. A demonstration of this via the detection of the Crab Pulsar from off-axis CZTI observations has been presented by (Anusree et al. 2021).
  
  The spectra of pulsars detected by Fermi-LAT in the $\gamma$-ray band (100~MeV-300 GeV) are usually described by a power-law with a high energy cutoff, with the power-law photon indices lying in the range 0 to 2.04 (2PC Catalog, Abdo et al. 2013).  To understand whether the underlying non-thermal emission continues to lower energies, determining the spectral properties at hard X-rays is crucial. While we can recover hard X-ray profiles of pulsars in CZTI, detailed spectroscopic analysis is rendered difficult by the strong background in the open detector.   
  
    A crude estimate of the spectral slope may, however, be obtained from the hardness ratio, namely the ratio of counts in two chosen energy bands. A simple relation\footnote{https://asd.gsfc.nasa.gov/Craig.Markwardt//bat-cal/hardness-ratio/} between the hardness ratio and power-law photon index has been derived for the Swift-BAT instrument, using its pre-launch response matrix. We develop this idea to establish a similar relation for off-axis CZTI observations, and apply it to the hard X-ray pulsars detected by the instrument. We utilise  instrument responses derived from AstroSat mass model simulations (Mate et al. 2021) in the spectral fitting package XSPEC (Arnaud 1996) to predict photon index values from hardness ratios measured in CZTI.  

This work also uses the new calibration results of the low-gain pixels in CZTI, discussed in Chattopadhyay et al (2021). This extends the detection of pulsars in CZTI into sub-MeV bands. The Crab pulse profile has a well-characterized energy dependence (Kuiper et al. 2001; Tuo et al. 2019). We obtain the Crab pulse profile up to 1~MeV from the low-gain pixels. To verify the new gain values of these pixels, we show that the observed ratios of the counts at the two pulse peaks and at the bridge are consistent with the literature.
 
    This paper is structured as follows. Section \ref{sec2} and \ref{methodplpred} describe the data analysis and methodology. Following a discussion of results in section \ref{sec4}, which includes validating the new gain values of low-gain pixels and the extension of detection in CZTI up to $\sim$1~MeV, we verify the photon index prediction from hardness ratios using the Crab pulsar. We also estimate with $\sim$80\% accuracy the photon indices of canonical soft $\gamma$-ray pulsars PSRJ1513-5908 and PSR J0835-4510. Finally, we show that the estimation  of photon index can be extended up to $\sim$1~MeV energies with proper calibration of low-gain pixels.


\section{Data \& Background }\label{sec2}
    This work is based on data from AstroSat-CZTI pointing observations released for public use on or before 30th July 2020. The rotational parameters of the pulsars under study have been obtained from the Timing programme pursued by the Fermi mission. A part of the work utilizes AstroSat Mass Model simulations. Information about the data and the analysis methods are described in this section.  Also discussed are the limitations and the scope of the method in the present context.
\subsection{AstroSat-CZTI} 
The CZTI instrument has four identical but independent quadrants. Each quadrant consists of sixteen 40 mm × 40 mm × 5 mm sized CZT detector modules, each further segmented into 256 pixels. A total of 16384 pixels results in a total collecting area of $\sim$1000~cm$^{2}$. The design and the function of the payload are described in detail in Bhalerao et al (2017) and Rao et al (2016). Primarily designed for imaging and spectroscopy up to 200~keV with $\sim 5^{\circ}$ field of view, the instrument becomes progressively transparent above $\sim$60~keV and acts as an all-sky open detector above $\sim$100~keV. Other than for detection of Pulsars as demonstrated in  (Anusree et al. 2021), this property of the instrument has also been utilized for the detection of many Gamma Ray Bursts (Rao et al. 2016; Chattopadhyay et al. 2019). 

\subsection{Detection of Hard X-ray Pulsars}
  While CZTI makes pointing observations of a target in the main field of view, photons from hard X-ray sources present elsewhere in the sky penetrate through the walls and are recorded in the detector. This way, since the launch of AstroSat, hard X-ray data has been continuously accumulating in the CZTI archive for more than half a decade. Photons from a pulsar arrive at the detector, modulated by its spin period.  Availability of precise timing ephemeris makes it possible to recover the pulse profiles from the CZTI archival data. We use the ephemerides of LAT pulsars generated by Fermi Timing programme (Ray et al. 2011; Kerr et al. 2015; Clark et al.2017) in conjunction with the Tempo2 software package (Hobbs et al. 2006) to generate polynomial coefficients (Polycos) describing the timing model of each pulsar. We have developed an algorithm to fold both $\gamma$-ray and X-ray events using the Solar System Barycentre Polycos. This algorithm can be used for data with gaps from any X-ray mission. More details of this method and its validation are described in Anusree et al (2021). We use the same method for obtaining all the pulse profiles in this work. The spin ephemeris of pulsars derived from Fermi observations and used in this work are listed in Tables \ref{table1} and \ref{table2}  respectively for PSR J0534+2200 and PSRJ1513-5908, along with the reference epochs.
 \begin{table}[]
 \tabularfont
 \caption{Fermi-LAT $\gamma$-ray timing solution of PSR J0534+2200 by Fermi Timing Observers. The numbers in parentheses
 denote $1\sigma$ errors in the parameters}
 \label{table1}. 
 \begin{tabular}{cc}
 \hline
 Parameters&   ~\\\midline
 Right ascension,~~ $\alpha$ &05:34:31.94\\
 Declination,~~$\delta$ &+22:00:52.1\\
 Valid MJD range  &54686-58767\\\\
 Pulse frequency,~~  $f~(s^{-1}$   &29.7169027333\\
 ~~                                 &(0.0000148379)\\
 First derivative , $ \dot{f}~(10^{-10} s^{-2} )$ &-3.71184342371 \\
 ~~                                  &(4.6591493867e-17)\\
 Second derivative ,$ \ddot{f}~(10^{-20} s^{-3} )$ &3.3226958153\\
 ~~                                       &(1.1931513864e-24)\\
 Third derivative , $ \dddot{f}~(10^{-28} s^{-4} )$ &1.2860657170\\
 ~~                                        &(9.2842295751e-32)\\
 PEPOCH        &55555 \\                     
 POSEPOCH       &50739\\                     
 TZRMJD &56730.15526586924  \\
 Solar system ephemeris model&DE405\\
 Time system &TDB\\
 \hline
 \end{tabular}
 \tablenotes{}
 \end{table}
\begin{table}[]
 \tabularfont
 \caption{Fermi-LAT $\gamma$-ray timing solution of PSR J1513-5908 by Fermi Timing Observers. The numbers in parentheses
 denote $1\sigma$ errors in the parameters}
 \label{table2}. 
 \begin{tabular}{cc}
 \hline
 Parameters&   ~\\\midline
 Right ascension,~~ $\alpha$ &15:13:55.62\\
 Declination,~~$\delta$ &-59:08:09.0  \\
 Valid MJD range  &54240-58647\\\\
 Pulse frequency,~~  $f~(s^{-1}$   &6.5970919120\\
 ~~                                 &(0.0000000001)\\
 First derivative , $ \dot{f}~(10^{-10} s^{-2} )$ &-0.6653060108 \\
 ~~                                  &(1.8154387575e-18)\\
 Second derivative ,$ \ddot{f}~(10^{-20} s^{-3} )$ &0.1894558542\\
 ~~                                       &(1.1931513864e-24)\\
 
 PEPOCH        &55336 \\                     
 POSEPOCH       &54710\\                     
 TZRMJD &55794.106598751587281       \\
 Solar system ephemeris model&DE405\\
 Time system &TDB\\
 \hline
 \end{tabular}
 \tablenotes{}
 \end{table}

\subsection{Characterization of low-gain pixels}
\begin{figure*}[!h]
\centering
\begin{subfigure}[b]{0.53\textwidth}
                \includegraphics[width=1.0\columnwidth,height=1.0\columnwidth,angle=0]{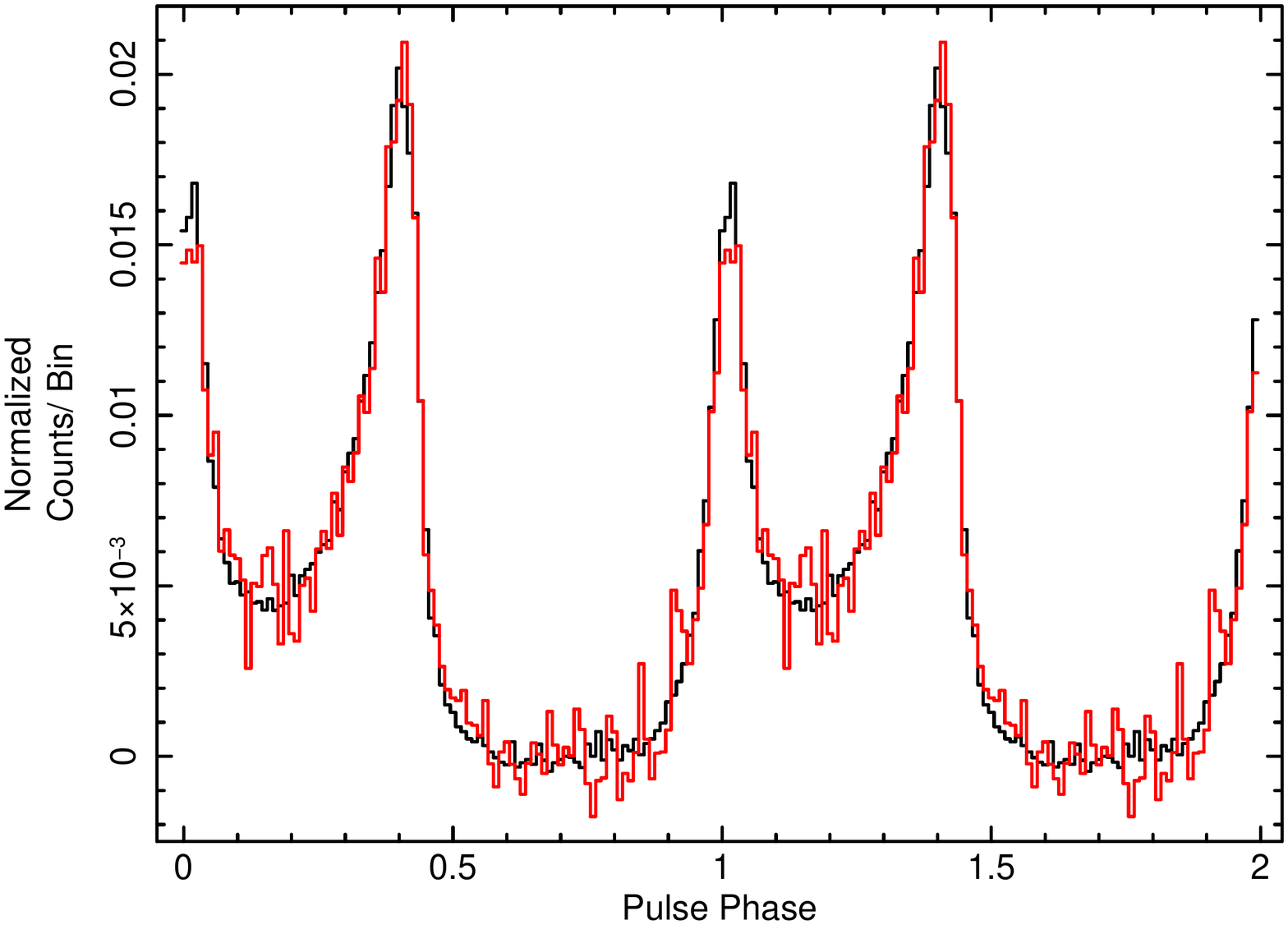}
               \caption{Mode of Detection : On-axis, Channels :150-512\\
               Profile: Black (normal pixels), Red (low-gain pixels)}
                \label{fig1(a)}
        \end{subfigure}%
        \begin{subfigure}[b]{0.53\textwidth}
                \includegraphics[width=1.0\columnwidth,height=1.0\columnwidth,angle=0]{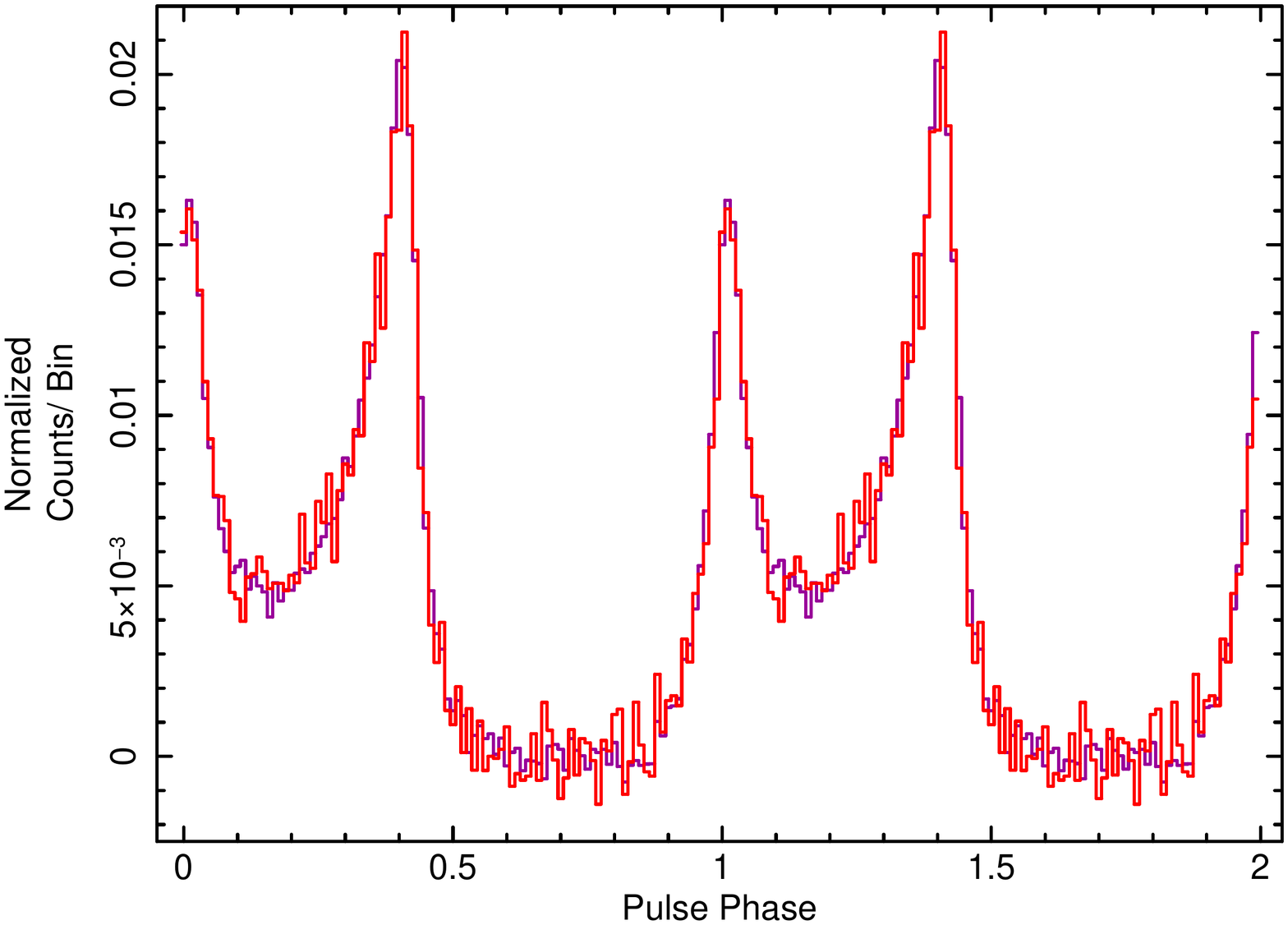}
                \caption{Mode of Detection : Off-axis, Channels :150-512 \\
                Profile: Purple (normal pixels), Red (low-gain pixels)}
                \label{fig1(b)}
        \end{subfigure}%

\caption{ According to Kuiper et al (2001), P2/P1 and Bridge/P1 ratio of Crab pulsar increase with the increase in energy up to $\sim$1MeV. In CZTI on-axis, P1 and P2 compared to normal pixels is seen diminishing and growing respectively, resulting in an increased P2/P1 ratio ($\sim$1.38 in good pixels and $\sim$1.50 in low-gain pixels) indicating higher energies of low-gain pixels. In the off-axis case, the pulse profile from both type of pixels are rather similar and can be attributed to the reduced sensitivity at lower energies. However, a marginal rise in P2 is observed. The results support the postulate by Chattopadhyay et al (2021) that the low-gain pixels have higher energy thresholds than the normal pixels.}
\label{fig1}
\end{figure*}

Soon after the launch of AstroSat, nearly 20\% of the CZTI pixels seemed to have an inadequate spectroscopic response (lower gain than expected). These pixels had since been excluded from scientific analysis. However, a recent re-calibration of these pixels has quantified their energy response and extended the energy range of CZTI spectra up to $\sim$1~MeV (Chattopadhyay et al. 2021). As a preliminary experiment, we obtained the Crab pulse profiles using 0.5~Ms of on-axis observations and 10~Ms of archival data where the Crab pulsar is within 5--70 degrees away from the pointing axis (hereafter off-axis observations). This was done using both normal pixels and low-gain pixels. The resulting high signal to noise ratio pulse profiles are shown in Figures \ref{fig1(a)} and \ref{fig1(b)} respectively.


One way to verify the new gains of these pixels is by looking at the energy dependent morphology of the Crab pulse profile. The profile consists of two pulses P1 and P2, joined by a ``Bridge''. In this work, we show that from low-gain pixels alone, pulse profiles of Crab can be obtained in CZTI sub-bands up to $\sim$1~MeV and the observed P2/P1 and Bridge/P1 ratios in these bands are consistent with the literature (section \ref{crabprofratios}). With these results, we show that we can extend the study of pulsars up to sub-MeV energies by including these pixels. 

\subsection{Response generation from AstroSat Mass Model outputs}\label{mmrsp}
A key aspect of spectroscopy with an X-ray instrument such as the CZTI is estimating the Redistribution Matrix (RMF) and the Auxiliary Response (ARF) of the detector. We model the energy response (RMF) of each pixel as a Gaussian spread about the incident energy, with the standard deviation determined from ground calibration.  We estimate the effective area from  GEANT4 simulation of the Astrosat mass model (Mate et al. 2020), in which monochromatic photons at different energies between 20 keV and 2 MeV are incident and the effective area as a function of both incident and detected energies is obtained to generate the ARF. We then estimate the pixel wise detector response by multiplying the RMF and the ARF. We apply a pixel mask, obtained from in-flight performance, to include the desired pixels. Our previous work had been carried out by including only the normal gain pixels at this stage, while in the current work we extend the same to include the low gain pixels as well. To the mass model simulation results, we apply appropriate Lower and Upper Level Discriminators (LLD and ULD) to replicate the actual pixel behaviour. We then generate the quadrant wise response by summing the responses of the appropriate pixels.

\section{Methodology for Estimation of Spectral Index}\label{methodplpred}

A Swift-BAT Calibration Note by C. Markwardt and D. Hullinger \footnote{https://asd.gsfc.nasa.gov/Craig.Markwardt//bat-cal/hardness-ratio/} demonstrates that a reasonable estimate of the spectrum of a Gamma-Ray Burst can be obtained using the hardness ratio, computed as the ratio of counts in one energy band to that in another, if the incident spectrum is a power-law in the bands of interest.  A simple relation between the hardness ratio and the power-law index is presented in the note.  Here we extend this method to the CZTI and obtain a calibration relation between the hardness ratio and photon index of pulsars in the CZTI band (80--200~keV for off-axis observations with normal gain pixels). We use the Crab pulsar as a test source.
\\
 We generate the instrument responses as described in section \ref{mmrsp} and use them  to find a simple relation between the hardness ratio and the power-law photon index. In the scenario of multiple directions, we generate an exposure-weighted response by the weighted addition of responses using Heasoft-FTOOL \textit{addrmf}. After generating the response matrix, we use the X-ray spectral fitting software (XSPEC, Arnaud 1996) to simulate power-law spectra with photon indices ranging from 0.5 to 3.0, for a hypothetical count rate of 1 and exposure 10ks. A statistically significant ratio of counts in 80.0-100.0 to 100.0-200.0 band is estimated from 100 such fake spectra. Then we define photon index as a function of Hardness Ratio by obtaining a simple relation between the two. This relation is then used to predict the photon index of pulsars in CZTI by measuring hardness ratios from observed pulse profiles.

\section{Results and Discussion}\label{sec4}
This section demonstrates the use of low-gain pixels to study Pulsars in CZTI and validate the new gains from the re-calibration. We also present the power-law indices we calculated using the method described in section 3.

\subsection{Crab Pulse Profiles up to $\sim$1~MeV in low-gain pixels }\label{crabprofratios}
    The energy dependence of the Crab pulse profile has been well known for a long time. Several models have been constructed over the years in an attempt to explain the behaviour (e.g.\ Cheng et al. 2000; Takata et al. 2007; Tang et al. 2008; Bai and Spitkovsky 2010; P\'{e}tri 2012; Philippov and Spitkovsky 2018), via a combination of magnetospheric structure and the position-dependent energy distribution of electron-positron pairs. However a fully self-consistent picture from first principles is yet to be developed. We use the observationally well characterized profile changes of the Crab pulsar with energy to assess the low-gain pixels in CZTI.

\begin{figure*}[]
\centering
\begin{subfigure}[b]{0.5\textwidth}
                \centering
                \includegraphics[width=1.0\columnwidth,height=1.45\columnwidth,angle=0]{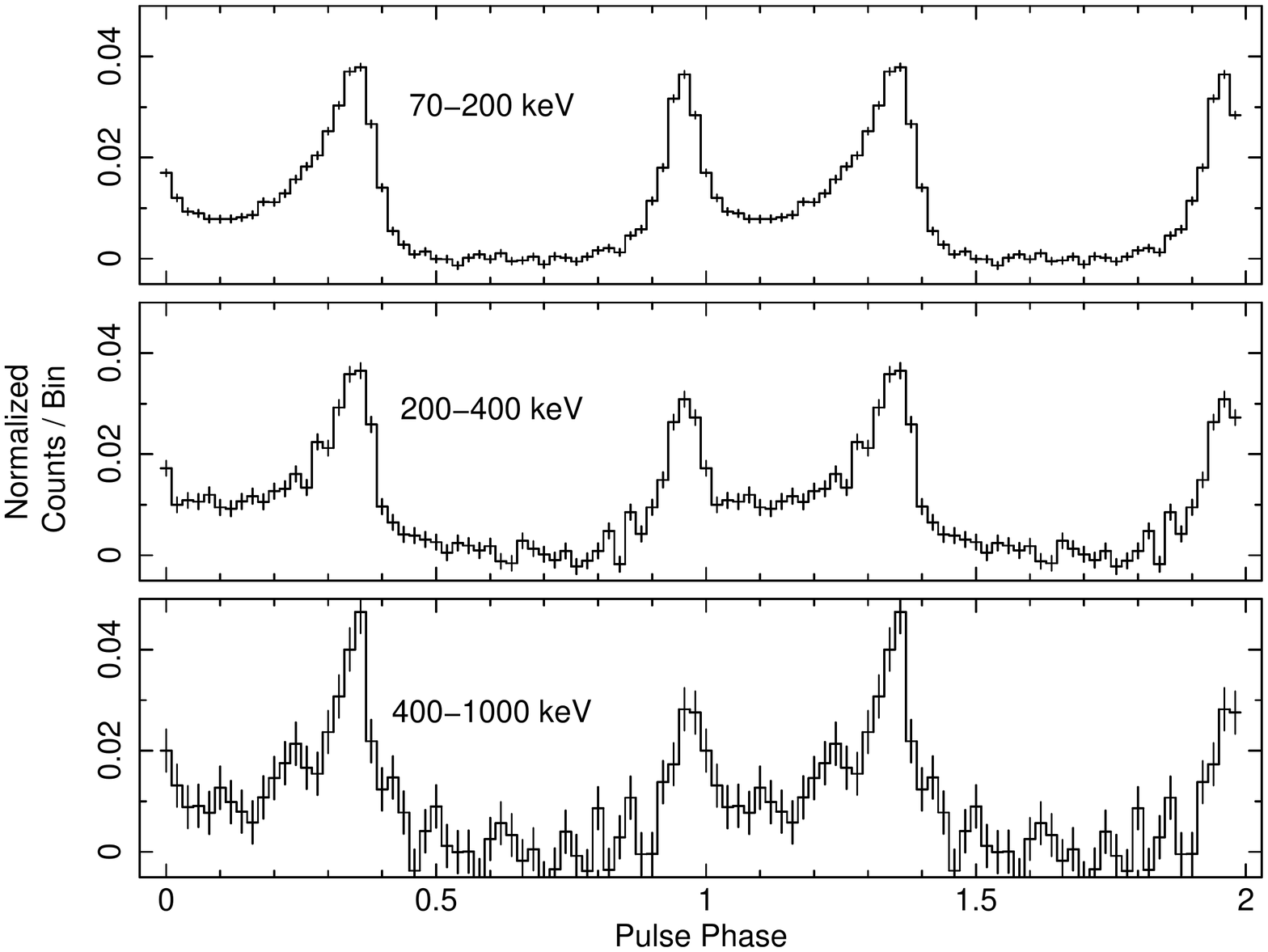}
               \caption{Mode of Detection: On-axis}
                \label{fig2(a)}
        \end{subfigure}%
        \begin{subfigure}[b]{0.5\textwidth}
                \centering
                \includegraphics[width=1.0\columnwidth,height=1.45\columnwidth,angle=0]{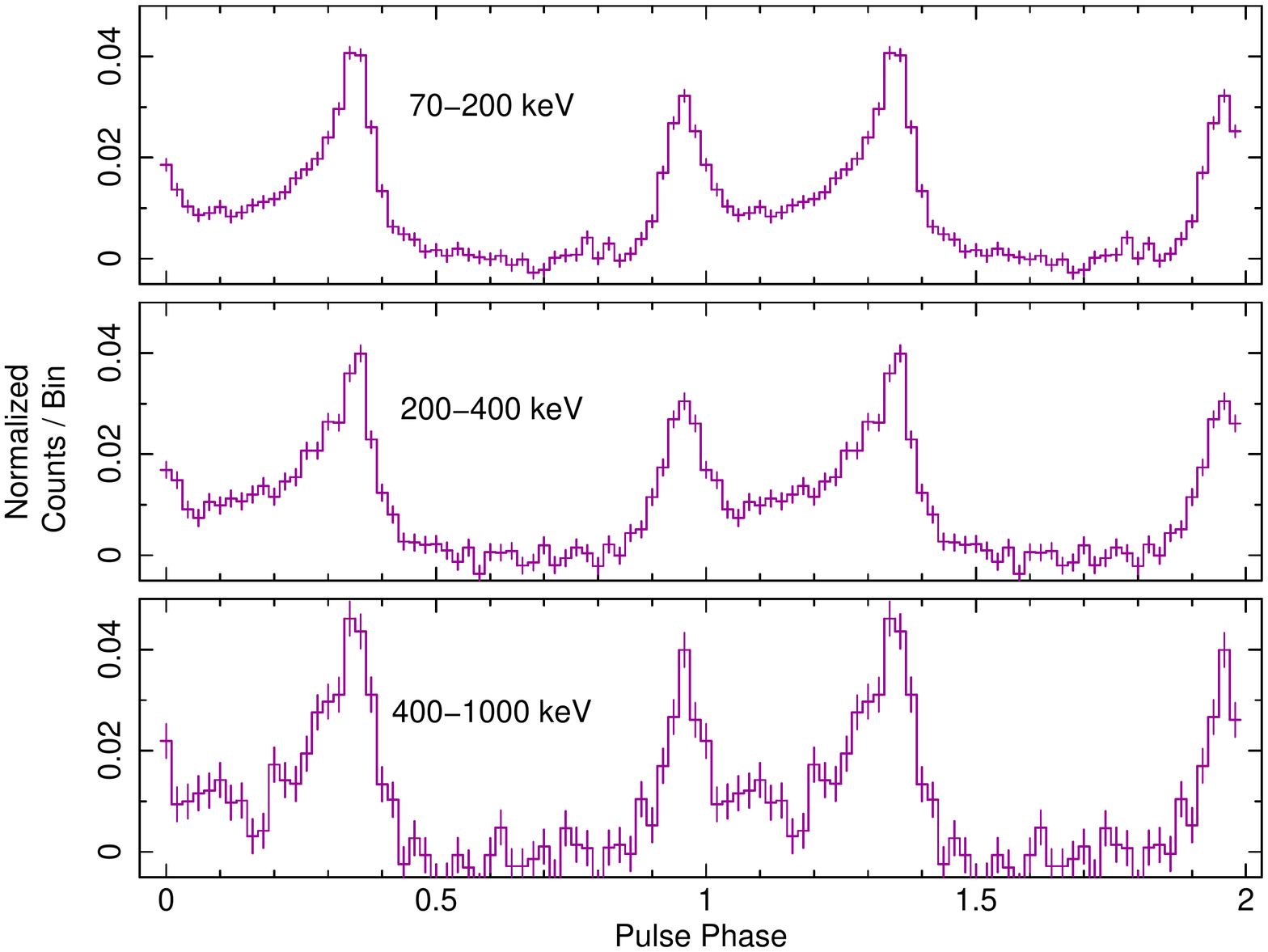}
                \caption{Mode of Detection: Off-axis}
                \label{fig2(b)}
        \end{subfigure}%

\caption{The energy-resolved pulse profile of Crab pulsar in low-gain pixels of all the CZTI quadrants after the gain correction. The increase in P2 and the bridge emission intensities relative to that of P1 is evident from the On-axis profiles. A more quantitative discussion of the profile shape, done towards verifying low-gain pixel gains, is available in Tables~\ref{table3} and \ref{table4} }
\label{fig2}
\centering
\end{figure*}

Obtained from the same channels,  Figure~\ref{fig1} shows Crab pulse profiles of higher P2/P1 ratio in low-gain pixels compared to good pixels, indicating that that the low-gain pixels respond to higher energies. For a more quantitative  verification of the new gains of these pixels obtained from the re-calibration, we applied it on 20 on-axis and 368 off-axis observations of Crab to create energy-resolved pulse profiles. In Figure~\ref{fig2}, we present energy-resolved Crab pulse profiles for low-gain pixels up to 1~MeV. 

    Kuiper et al. (2001) provide a detailed characterization of the pulse profile of the Crab pulsar over the 0.75–30 MeV band. The P2 and the Bridge emission intensity increases with increasing energy up to $\sim$1~MeV and above $\sim$1~MeV, P1 dominates P2, and the bridge emission becomes less significant. In Table~\ref{table4}, we report the P2/P1 ratio and the Bridge/P1 ratio in low-gain pixels obtained by adopting the phase definitions used by Kuiper et al. (2001) (P1: 0.94-1.04, P2: 0.32-0.43, Bridge/Ip: 0.14-0.25, Off Pulse: 0.52-0.88). Increasing the integration to 1 MeV resulted in an increased P2/P1 and Bridge/P1 ratio, agreeing well with the trend seen in Kuiper et al (2001).

A more recent study by Tuo et al. (2019) characterizes the energy dependence of Crab profile in the 15--250~keV band using the Insight-HXMT mission and present a lower limit of 1 for the P2/P1 ratio. Adopting their definitions of peaks, with P1 and P2 as the local maximum of the leading and trailing peak respectively, we compute the ratio in the 100--1000~keV band in CZTI low-gain pixels. The ratios thus obtained in CZTI as reported in Table~\ref{table3} are well above the lower limit of 1.

We obtained the energy-resolved Crab pulse profile from low-gain pixels, as shown in Figure.\ref{fig2} which, together with the observed profile characteristics reported in Tables~\ref{table3} and \ref{table4} verify the higher energy thresholds of low-gain pixels in CZTI and validates the new gains from re-calibration. We also note that the P2/P1 ratio is more prominent in the off-axis data than in the on-axis, consistent with the increased sensitivity to high energy photons in the off-axis scenario.

\begin{table*}[]
\centering
\tabularfont
\caption{ P2/P1 ratio of Crab Profile from low-gain pixels in CZTI compared with corresponding  lower limit by Insight-HXMT (Tuo et al. 2019)}
\label{table3}
\begin{tabular}{lcccc}
\topline
E-band& \textbf{Mode of Detection}&\textbf{P2/P1 (this work))}& \textbf{P2/P1 (Insight-HXMT)}&\textbf{S/N}\\
\hline
100-1000~keV &On-axis&1.15$\pm$0.04&$>$1&58.6\\
\hline
 100-1000~keV&Off-axis&1.32$\pm$0.06&$>$1&45.0\\
 \hline
\end{tabular}
\\
P1 and P2 here are the local maxima of Peak1 and Peak2 and S/N is the signal to noise ratio of the pulse profile in CZTI.
\end{table*}
\begin{table*}[]
\centering
\tabularfont
\caption{P2/P1 ratio and Bridge/P1 ratio of Crab Profile from low-gain pixels in CZTI compared with literature.
}
\label{table4}
\begin{tabular}{lcccccc}
\topline

E-band& \textbf{Mode of Detection}&\textbf{P2/P1(this work)}&\textbf{P2/P1$^{*}$}& \textbf{Bridge/P1(this work)}& \textbf{Bridge/P1 $^{*}$}&\textbf{S/N}\\
\hline
100-200~keV&On-axis&1.38$\pm$0.07&1.4-1.9&0.41$\pm$0.04&0.4-0.7&42.6\\

100-1000~keV&On-axis&1.46$\pm$0.04&1.4-1.9&0.51$\pm$0.04&0.4-0.7&39.9\\

\hline
100-200~keV&Off-axis&1.58$\pm$0.07&1.4-1.9&0.52$\pm$0.06&0.4-0.7&23.1\\

100-1000~keV&Off-axis&1.66$\pm$0.06&1.4-1.9&0.65$\pm$0.04&0.4-0.7&29.7\\
\hline
\end{tabular}
Here, P2/P1 and Bridge/P1 are the intensity ratios of the profile features estimated from the background-subtracted counts in respective phase intervals. S/N is the signal to noise ratio of the pulse profile in respective CZTI bands.\\
$^{*}$ The range in which P2/P1 and Bridge/P1 ratios increase as energy is increased from 100~keV to 1~MeV, as observed by Kuiper et al (2001).  
\end{table*}

\subsection{Photon Index from hardness ratios}\label{plprediction}
In order to verify the method, we estimated the power-law index for our test source Crab as described in Section~\ref{methodplpred}. The results thus obtained are presented in this section for validation. We also present our results of power-law index prediction for the pulsar PSR J1513-5908.

\subsubsection{Phase Resolved Spectra of Crab:} \label{phaseresolved}  The total pulsed spectrum of Crab is a broken power-law in the 60-200 keV band, with power-law indices 1.85 and 2.20 with a break around $\sim$130keV (Ulmer et al. 1995); hence can not be estimated by our method. However, the phase-resolved spectrum of Crab is consistent with a simple power law. In a study using INTEGRAL data, power-law fits to the spectrum of selected phase intervals of Crab profile in the 80-200 keV yield photon indices 2.30 ±0.05, 1.98 ±0.10, and 2.14 ±0.05, respectively, for the phase intervals that define  P1, Bridge/Ip and P2 (Massaro et al. 2006).

\begin{figure*}[!h] 
\centering
  \begin{subfigure}[b]{0.5\linewidth}
    \centering
    \includegraphics[width=8cm,height=7cm]{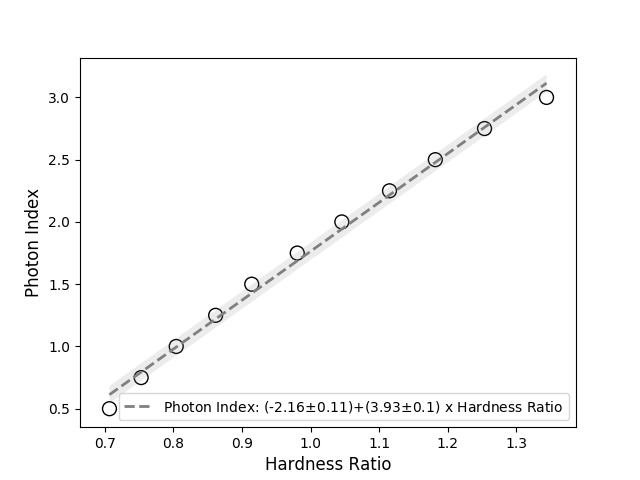} 
    \caption{Hardness Ratio = $\frac{{\rm counts~in}\; 80-100\; {\rm keV}}{{\rm counts~in}\; 100-130\; {\rm keV}}$} 
    \label{fig3:a} 
    \vspace{1ex}
  \end{subfigure}
  \begin{subfigure}[b]{0.5\linewidth}
   \centering
    \includegraphics[width=8cm,height=7cm]{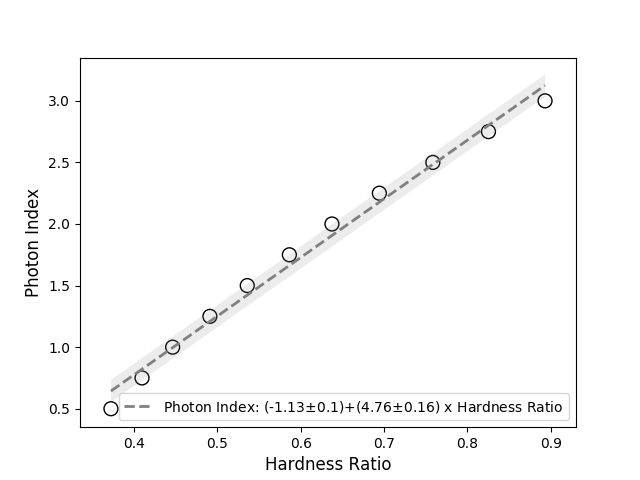} 
    \caption{Hardness Ratio = $\frac{{\rm counts~in}\; 80-100\; {\rm keV}}{{\rm counts~in}\; 100-160\; {\rm keV}}$} 

    \label{fig3:b} 
    \vspace{1ex}
  \end{subfigure} 
  \begin{subfigure}[b]{0.5\linewidth}
    \centering
    \includegraphics[width=8cm,height=7cm]{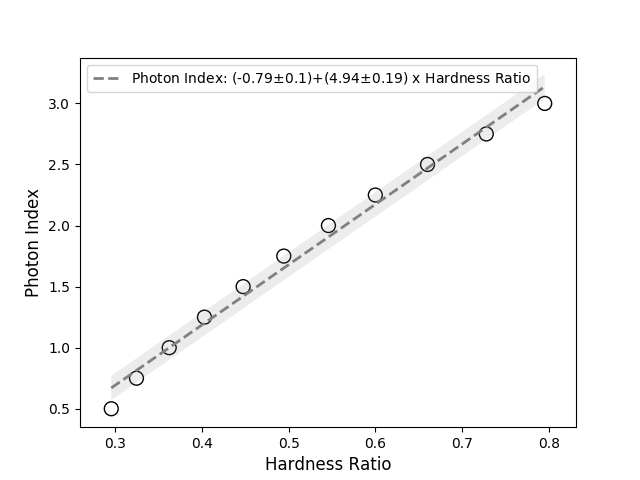} 
     \caption{Hardness Ratio = $\frac{{\rm counts~in}\; 80-100\; {\rm keV}}{{\rm counts~in}\; 100-180\; {\rm keV}}$} 

    \label{fig3:c} 
  \end{subfigure}
  \begin{subfigure}[b]{0.5\linewidth}
   \centering
    \includegraphics[width=8cm,height=7cm]{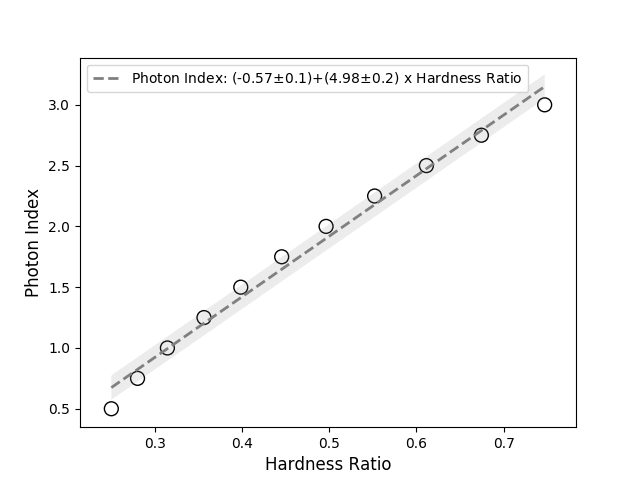} 
    \caption{Hardness Ratio = $\frac{{\rm counts~in}\; 80-100\; {\rm keV}}{{\rm counts~in}\; 100-200\; {\rm keV}}$} 

    \label{fig3:d} 
  \end{subfigure} 
  \caption{Straight line fits to the variation of power-law photon indices with  hardness ratios in good pixels.  The hardness ratios are obtained from simple power-laws simulated using the instrument responses derived from AstroSat mass model (Mate et al. 2021) in the spectral fitting package XSPEC (Arnaud 1996).  The relations presented here are used to predict the photon indices in Table \ref{table5}}
  \label{fig3} 
\end{figure*}

For $\sim$1000~ks CZTI observations of Crab pulsar, following the procedure described in Section~\ref{methodplpred}, we obtained relations between power-law index and hardness ratios  defined by the ratio of counts in the 80.0--100.0~keV band to those in 100.0--130.0~keV, 100.0--160.0~keV, 100.0--180.0~keV and 100.0--200.0~keV respectively. In each case, as shown in Figure~\ref{fig3}, the relation was described by a straight line. We then used these relationships to predict the power-law index describing the phase-resolved spectrum of Crab. The power-law indices predicted from the hardness ratios observed in the phase intervals for P1, P2, and Bridge defined by Massaro et al (2006) (P1: 0.98-1.01, P2: 0.38-0.41, Bridge/Ip: 0.08-0.27,  Off Pulse: 0.60-0.80) presented in Table \ref{table5} is $\sim$90\% accurate and verifies our method. 

\begin{table*}[!h]
\centering
 \caption{Phase Resolved Spectrum of Crab in CZTI Normal Pixels}
 \begin{tabular}{lcccccc}
 \label{table5}
Profile Feature of Crab& $^{a}$80-100~keV& $^{b}$100-130~keV&$^{c}$ Hardness Ratio & $^{d}$PL (Pred.) & $^{e}$ PL (Ref.)\\
\hline
P1&136961$\pm$2450&121623$\pm$3230&1.13$\pm$0.040&2.27$\pm$0.22&$2.30\pm 0.05$\\
P2&163395$\pm$2450&150454$\pm$3230&1.09$\pm$0.028&2.12$\pm$0.19&$2.14\pm 0.05$\\
Bridge/Ip &217869$\pm$2450&200605$\pm$3230&1.09$\pm$0.021&2.12$\pm$0.18&$1.98\pm 0.10$\\

\hline
& 80-100~keV&100-160~keV& & &\\
\hline
P1&136961$\pm$2450&183735$\pm$5145&0.75$\pm$0.025&2.44$\pm$0.20&$2.30\pm 0.05$\\
P2&163395$\pm$2450&232691$\pm$5145&0.70$\pm$0.019&2.20$\pm$0.18&$2.14\pm 0.05$\\
Bridge/Ip &217869$\pm$2450&311160$\pm$5145&0.70$\pm$0.014&2.20$\pm$0.17&$1.98\pm 0.10$\\
\hline
& 80-100~keV& 100-180~keV& &&\\
\hline
P1&136961$\pm$2450&208870$\pm$5784&0.66$\pm$0.022&2.47$\pm$0.19&$2.30\pm 0.05$\\
P2&163395$\pm$2450&265015$\pm$5784&0.62$\pm$0.016&2.27$\pm$0.17&$2.14\pm 0.05$\\
Bridge/Ip &217869$\pm$2450&360941$\pm$5784&0.60$\pm$0.012&2.17$\pm$0.16&$1.98\pm 0.10$\\

\hline
&80-100~keV&100-200~keV&&&\\\hline
P1&136961$\pm$2450&222896$\pm$6287&0.61$\pm$0.021&2.47$\pm$0.19&$2.30\pm 0.05$\\
P2&163395$\pm$2450&285071$\pm$6287&0.57$\pm$0.015&2.27$\pm$0.17&$2.14\pm 0.05$\\
Bridge/Ip &217869$\pm$2450&388893$\pm$6287&0.56$\pm$0.011&2.22$\pm$0.16&$1.98\pm 0.10$\\
\hline
 \end{tabular}
 
 $^{a,b}$ Background subtracted counts in the CZTI bands specified and $^{c}$ is the ratio of the two \\
 $^{d}$  Power-Law index predicted from observed hardness ratios in CZTI\\
$^{e}$ Power-Law index obtained using  \textit {INTEGRAL } data in 80-200~keV band  (Massaro et al. 2006)
 
 \end{table*}
 
\subsubsection{Photon index of PSR~J1513-5908:}

PSR~J1513-5908, a canonical soft $\gamma$-ray pulsar, is among the 18 hard X-ray pulsars detected in CZTI off-axis data to date (K.G Anusree et al. 2022b, In Prep.). The total pulsed spectrum of J1513 in the 94.0--240.0~keV band is well-fit with a simple power-law of photon index of 1.64$\pm$0.40 (Gunji et al. 1994). For a better comparison with the known spectrum, we obtained CZTI pulse profiles in 94--100~keV and 100-200~keV as shown in Figure \ref{fig4}, using archival data spanning $\sim$3 years where the pulsar is within 5--70 degrees of the pointing. In Figure~\ref{fig5}, we present the relationship connecting power-law index and hardness ratio defined as the ratio of counts in 94.0--100.0~keV band to those in 100.0--200.0~keV band, obtained by the method described in section \ref{methodplpred}. The photon index of $1.64\pm0.58$ reported in Table.\ref{table6}, predicted from the hardness ratio measured from the profiles in Figure~\ref{fig5} is consistent with the results of (Gunji et al. 1994).

\begin{figure*}[!htbp]
\centering
\begin{subfigure}[b]{0.53\textwidth}
        \centering
        \includegraphics[width=1.0\columnwidth,height=1.0\columnwidth,angle=0]{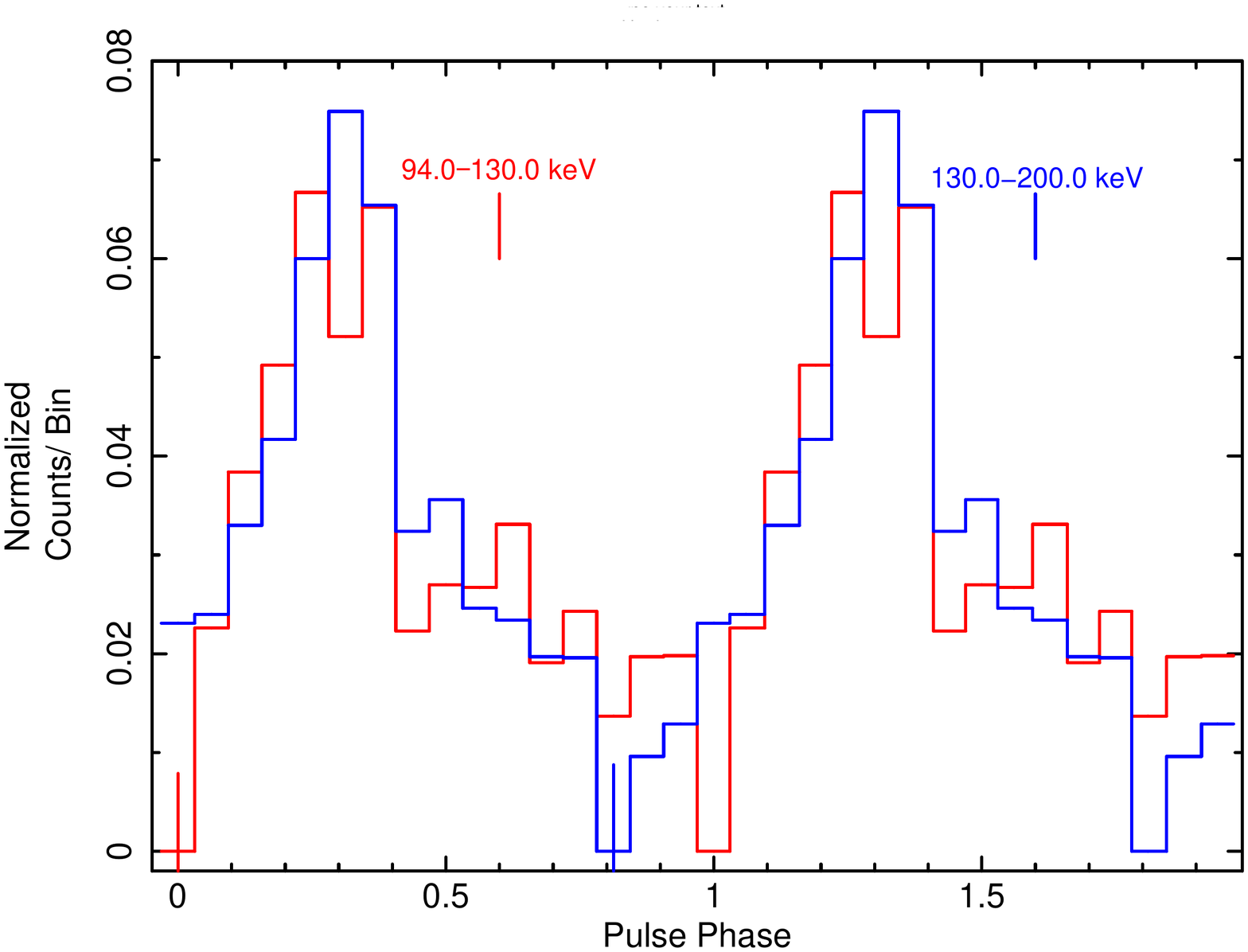}
        \label{fig4:a} 
        \caption{}
        \label{fig4}
\end{subfigure}%
\begin{subfigure}[b]{0.53\textwidth}
        \centering
        \includegraphics[width=1.0\columnwidth,height=.98\columnwidth,angle=0]{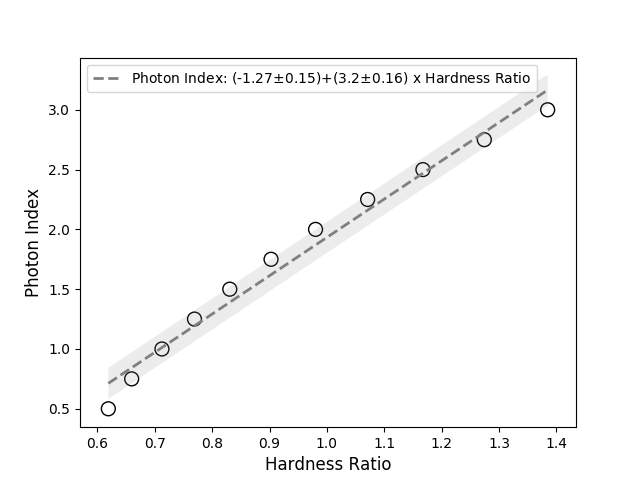}
        \label{fig4:b} 
        \caption{Hardness Ratio = $\frac{{\rm counts~in}\; 94-130\; {\rm keV}}{{\rm counts~in}\; 130-200\; {\rm keV}}$} 
        \label{fig4}
\end{subfigure}%

\caption{The 94--130~keV  \& 130--200~keV pulse profiles of PSRJ1513-5908  in good pixels of all the CZTI quadrants. The photon index predicted from the hardness ratios obtained from these profiles are shown in Table~\ref{table6}.}
\end{figure*}

\begin{table*}[!htbp]
\centering
\caption{Total Pulse Spectrum of PSR J1513-5908 in CZTI Good Pixels}
\begin{tabular}{lccccc}
Pulsar Name& $^{a}$94-130~keV& $^{b}$130-200~keV& $^{c}$Hardness Ratio & $^{d}$PL (Pred.) & $^{e}$ PL (Ref.)\\
\hline
PSRJ1513-5908 &824904$\pm$71525&909280$\pm$83324&0.91$\pm$0.11&1.64$\pm$0.41&$1.64\pm0.40$\\
\hline
\end{tabular}
\label{table6}
\\
$^{a,b}$ Background subtracted counts in CZTI bands specified and $^{c}$ is the ratio of two\\
 $^{d}$  Power-Law index predicted from observed hardness ratios in CZTI\\
$^{e}$ Power-Law index obtained using  \textit {WELCOME} data in 94-240~keV band  (Gunji et al. 1994)
\end{table*}

\subsection{Photon Index prediction from low-gain pixels}
It is clear from Section~\ref{crabprofratios} that the inclusion of the low-gain pixels extends the energy band of CZTI up to $\sim$1~MeV. We repeated the power-law index prediction for phase-resolved Crab spectra similar to that in Section~\ref{phaseresolved} but using the low-gain pixels. For this purpose, we generated response matrices from Mass Model using updated calibration files. We present the comparison of our results with the well-known spectrum of Crab in Table~\ref{table7}. With these results, we argue that with proper calibration, the inclusion of low-gain pixels can extend the power-law index prediction in CZTI up to $\sim$1~MeV.
\begin{figure*}[] 
\centering
  \begin{subfigure}[b]{0.5\linewidth}
    \centering
    \includegraphics[width=8cm,height=7cm]{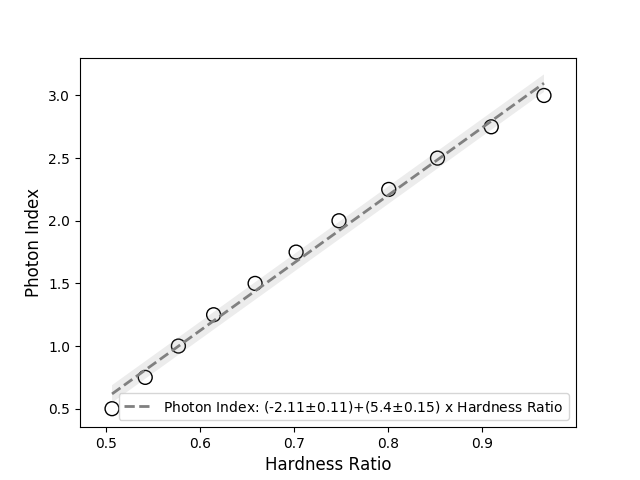} 
    \caption{Hardness Ratio = $\frac{{\rm counts~in}\; 80-100\; {\rm keV}}{{\rm counts~in}\; 100-130\; {\rm keV}}$} 
    \label{fig5:a} 
    \vspace{1ex}
  \end{subfigure}
  \begin{subfigure}[b]{0.5\linewidth}
   \centering
    \includegraphics[width=8cm,height=7cm]{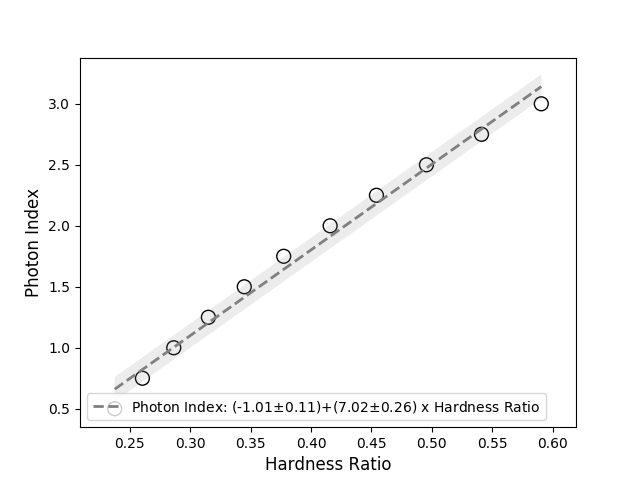} 
    \caption{Hardness Ratio = $\frac{{\rm counts~in}\; 80-100\; {\rm keV}}{{\rm counts~in}\; 100-160\; {\rm keV}}$} 

    \label{fig5:b} 
    \vspace{1ex}
  \end{subfigure} 
  \begin{subfigure}[b]{0.5\linewidth}
    \centering
    \includegraphics[width=8cm,height=7cm]{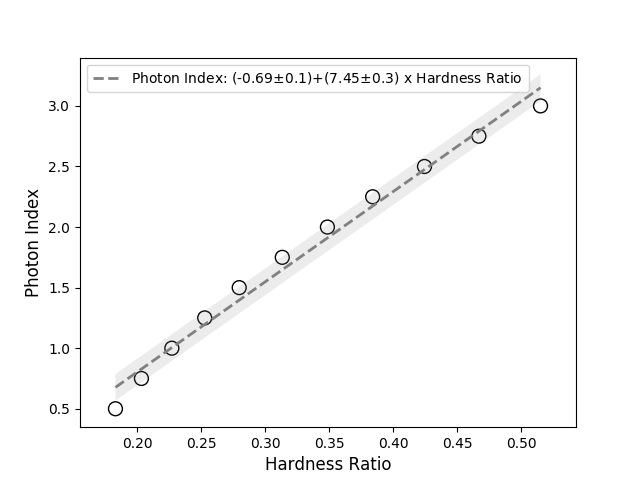} 
     \caption{Hardness Ratio = $\frac{{\rm counts~in}\; 80-100\; {\rm keV}}{{\rm counts~in}\; 100-180\; {\rm keV}}$} 

    \label{fig5:c} 
  \end{subfigure}
  \begin{subfigure}[b]{0.5\linewidth}
   \centering
    \includegraphics[width=8cm,height=7cm]{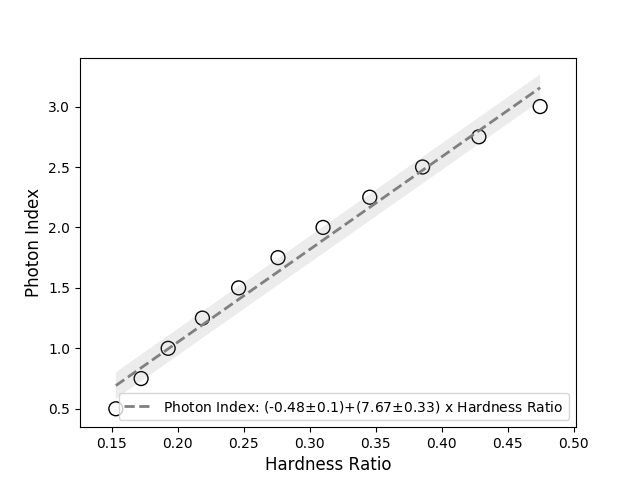} 
    \caption{Hardness Ratio = $\frac{{\rm counts~in}\; 80-100\; {\rm keV}}{{\rm counts~in}\; 100-200\; {\rm keV}}$} 

    \label{fig5:d} 
  \end{subfigure} 
  \caption{Straight line fits to the variation of hardness ratio with power-law photon index in low-gain pixels.  The hardness ratios are obtained from simple power-laws simulated using the instrument responses derived from AstroSat mass model (Mate et al. 2021) in the spectral fitting package XSPEC (Arnaud 1996). The relations presented here are used to predict the photon indices in Table~\ref{table7}.}
  \label{fig5} 
\end{figure*}
\begin{table*}[]
\centering
 \caption{Phase Resolved Spectrum of Crab in CZTI Low-Gain Pixels }
 \begin{tabular}{lcccccc}
Profile Feature of Crab& $^{a}$80-100~keV& $^{b}$100-130~keV&$^{c}$ Hardness Ratio & $^{d}$PL (Pred.) & $^{e}$ PL (Ref.)\\
\hline
P1 &18146$\pm$1128&22706$\pm$1176&0.80$\pm$0.060&2.21$\pm$0.36&$2.30\pm 0.05$\\
P2& 20336$\pm$1128&26004$\pm$1176&0.78$\pm$0.060&2.10$\pm$0.36&$2.14\pm 0.05$\\
Bridge&24565$\pm$1128&33012$\pm$1176&0.74$\pm$0.040&1.88$\pm$0.27&$1.98\pm 0.10$\\

\hline
& 80-100&100-160~keV& & &\\
\hline
P1 &18146$\pm$1128&38870$\pm$1678&0.47$\pm$0.035&2.29$\pm$0.30&$2.30\pm 0.05$\\
P2& 20336$\pm$1128&44391$\pm$1678&0.46$\pm$0.031&2.22$\pm$0.27&$2.14\pm 0.05$\\
Bridge&24565$\pm$1128&57839$\pm$1678&0.42$\pm$0.023&1.94$\pm$0.22&$1.98\pm 0.10$\\

\hline
& 80-100& 100-180~keV& &&\\
\hline
P1 &18146$\pm$1128&45000$\pm$1845&0.40$\pm$0.030&2.29$\pm$0.27&$2.30\pm 0.05$\\
P2& 20336$\pm$1128&53066$\pm$1845&0.38$\pm$0.025&2.14$\pm$0.24&$2.14\pm 0.05$\\
Bridge&24565$\pm$1128&67717$\pm$1845&0.36$\pm$0.019&1.99$\pm$0.21&$1.98\pm 0.10$\\

\hline
&80-100&100-200~keV&&&\\\hline
P1 &18146$\pm$1128&50175$\pm$1849&0.36$\pm$0.026&2.28$\pm$0.25&$2.30\pm 0.05$\\
P2& 20336$\pm$1128&60312$\pm$1849&0.34$\pm$0.021&2.12$\pm$0.22&$2.14\pm 0.05$\\
Bridge&24565$\pm$1128&76759$\pm$1849&0.32$\pm$0.017&1.97$\pm$0.20&$1.98\pm 0.10$\\
\hline
 \end{tabular}
 \\
 
 $^{a,b}$ Background subtracted counts  from low-gain pixels in the CZTI bands specified and $^{c}$ is the ratio of the two \\
 $^{d}$  Power-Law index predicted from observed hardness ratios in CZTI\\
$^{e}$ Power-Law index obtained using  \textit {INTEGRAL } data in 80-200~keV band  (Massaro et al. 2006)
\label{table7}

\end{table*} 

\section{Conclusion}
The CZTI instrument onboard AstroSat, an open detector above $\sim$100~keV, is a promising tool for studying the hard X-ray properties of pulsars. As reported earlier, a custom algorithm detects pulsars in the CZTI band from long integration of the archival data spanning $\sim$years (Anusree et al. 2021). In this work, we have extended this method to include an additional $\sim$20\% of the CZTI pixels that were excluded earlier due to their lower gain values. A recent re-calibration of these pixels demonstrated higher thresholds that extend the energy range in CZTI up to $\sim$1 MeV (Chattopadhyay et al. 2021). One way to verify the corrected gains of these pixels is to look at the pulse profiles of Crab in these pixels. In this work, we use the gains of low-gain pixels from re-calibration and recover energy-resolved pulse profiles of Crab up to $\sim$1~MeV. Also, we calculate from these pixels the P2/P1 and Bridge/P1 ratios of Crab pulsar at sub-MeV CZTI bands and verify that the ratios are consistent with the known values. Our results show that with proper calibration, the inclusion of these pixels can extend the detection of pulsars in CZTI up to $\sim$1~MeV.

Detailed spectroscopy of these pulsars over the full CZTI band is impossible due to the limited count rate. We have presented a successful method to predict the spectrum of pulsars detected in CZTI, assuming the radiation spectrum is a power law in the CZTI band. Using AstroSat Mass Model simulations, we calibrate the variation of hardness ratio with power-law index in CZTI. We then use it to predict the power-law index from the hardness ratios obtained from actual data. We present our results for the phase-resolved spectrum of PSR J0534+2200 and the total pulsed emission of PSR J1513-5908. The recovered photon indices are accurate to within $\sim 20$\%. The results are promising as this enables the comparison of the AstroSat-CZTI (80.0--200~keV) and Fermi-LAT (0.1--300~GeV) spectrum of LAT pulsars to check whether the hard X-ray and $\gamma$-ray emission have a common continuum. Inferences from this work will be used to obtain sub-MeV profiles and constrain the power-law spectral index of other pulsars using CZTI.

\section*{Acknowledgements}
We thank the anonymous referee for comments and suggestions that significantly improved the paper. This work uses data from the Indian astronomy mission AstroSat, archived at the Indian Space Science Data Centre (ISSDC). The instrument CZTI was built by a TIFR-led consortium of institutes across India, including VSSC, IUCAA, URSC, PRL, and SAC. The Indian Space Research Organisation funded, facilitated and managed the project. We extend our acknowledgement to the team members of the CZTI POC at IUCAA for helping with the aggregation of data. We also thank Fermi Timing Observers Paul Ray and Matthew Kerr for their timely and favorable response in providing LAT ephemeris for the pulsars.  We thank the  HPC facility at IUCAA, where we carried out all the data analysis. Anusree K. G. acknowledges support for this work from the DST-INSPIRE Fellowship grant, IF170239, under the Ministry of Science and Technology, India.


\balance

\end{document}